\documentclass[twocolumn,showpacs,pre,preprintnumbers,amsmath,amssymb,aps,floatfix]{revtex4}
\usepackage{graphicx}
\usepackage{dcolumn}
\usepackage{bm}
\usepackage{color}

\begin{document}

\title{Increased accuracy  of  ligand sensing by  receptor internalization}

\author{Gerardo Aquino}
\author{Robert G.  Endres }
\affiliation{Division of Molecular Biosciences $\&$ Centre for Integrated Systems Biology at Imperial College, Imperial College London,  SW7 2AZ, London, UK}
\date{\today }
\newcommand{\red}{\textcolor{red}}
\newcommand{\blue}{\textcolor{blue}}
\newcommand{\Sss}{\scriptscriptstyle}
\newcommand{\Ss}{\scriptstyle}
\newcommand{\D}{\dysplaystyle}
\newcommand{\T}{\textstyle}
\newcommand{\e}{{\rm e}}
\newcommand{\veps}{\varepsilon}
\newcommand{\epss}{\varepsilon_{\sigma}}
\newcommand{\epsv}{\varepsilon_{V}}
\newcommand{\lgl}{\langle}
 \newcommand{\rgl}{\rangle}
\newcommand{\Vh}[1]{\hat{#1}}
\newcommand{\Aa}{A^1_{\epsilon}}
\newcommand{\Ab}{A^{\epsilon}_L}
\newcommand{\Ae}{A_{\epsilon}}
\newcommand{\finn}[1]{\phi^{\pm}_{#1}}
\newcommand{\ea}{e^{-|\alpha|^2}}
\newcommand{\eb}{\frac{e^{-|\alpha|^2} |\alpha|^{2 n}}{n!}}
\newcommand{\ebbb}{\frac{e^{-3|\alpha|^2} |\alpha|^{2 (l+n+m)}}{l!m!n!}}
\newcommand{\ass}{\alpha}
\newcommand{\as}{\alpha^*}
\newcommand{\fb}{\bar{f}}
\newcommand{\gb}{\bar{g}}
\newcommand{\la}{\lambda}
 \newcommand{\sz}{\hat{s}_{z}}
\newcommand{\sy}{\hat{s}_y}
\newcommand{\sx}{\hat{s}_x}
\newcommand{\sio}{\hat{\sigma}_0}
\newcommand{\six}{\hat{\sigma}_x}
\newcommand{\siz}{\hat{\sigma}_{z}}
\newcommand{\siy}{\hat{\sigma}_y}
\newcommand{\vhsig}{\vec{\hat{\sigma}}}
\newcommand{\hsig}{\hat{\sigma}}
\newcommand{\hH}{\hat{H}}
\newcommand{\hU}{\hat{U}}
\newcommand{\hA}{\hat{A}}
\newcommand{\hB}{\hat{B}}
\newcommand{\hC}{\hat{C}}
\newcommand{\hD}{\hat{D}}
\newcommand{\hV}{\hat{V}}
\newcommand{\hW}{\hat{W}}
\newcommand{\hK}{\hat{K}}
\newcommand{\hX}{\hat{X}}
\newcommand{\hM}{\hat{M}}
\newcommand{\hN}{\hat{N}}
\newcommand{\te}{\theta}
\newcommand{\vze}{\vec{\zeta}}
\newcommand{\vet}{\vec{\eta}}
\newcommand{\vx}{\vec{\xi}}
\newcommand{\vc}{\vec{\chi}}
\newcommand{\hro}{\hat{\rho}}
\newcommand{\vro}{\vec{\rho}}
\newcommand{\hR}{\hat{R}}
\newcommand{\half}{\frac{1}{2}}
\renewcommand{\d}{{\rm d}}
\renewcommand{\top }{ t^{\prime } }
\newcommand{\oz}{{(0)}}
\newcommand{\sint}{{\rm si}}
\newcommand{\cint}{{\rm ci}}
\newcommand{\de}{\delta}
\newcommand{\ep}{\varepsilon}
\newcommand{\De}{\Delta}
\newcommand{\eps}{\varepsilon}
\newcommand{\si}{\hat{\sigma}}
\newcommand{\om}{\omega}
\newcommand{\tr}{{\rm tr}}
\newcommand{\ha}{\hat{a}}
\newcommand{\gam}{\gamma ^{(0)}}
\newcommand{\pe}{\prime}
\newcommand{\BEQ}{\begin{equation}}
\newcommand{\EEQ}{\end{equation}}
\newcommand{\BEN}{\begin{align}}
\newcommand{\EEGN}{\end{align}}
\newcommand{\BES}{\begin{subequations}}
\newcommand{\EES}{\end{subequations}}
\newcommand{\BEA}{\begin{eqnarray}}
\newcommand{\EEA}{\end{eqnarray}}
\newcommand{\sph}{spin-$\frac{1}{2}$ }
\newcommand{\ad}{\hat{a}^{\dagSger}}
\newcommand{\add}{\hat{a}}
\newcommand{\spp}{\hat{\sigma}_+}
\newcommand{\smm}{\hat{\sigma}_-}
\newcommand{\fin}[1]{|\phi^{\pm}_{#1}\rangle}
\newcommand{\finp}[1]{|\phi^{+}_{#1}\rDanah Matthewsangle}
\newcommand{\finm}[1]{|\phi^{-}_{#1}\rangle}
\newcommand{\lfin}[1]{\langle \phi^{\pm}_{#1}|}
\newcommand{\lfinp}[1]{\langle \phi^{+}_{#1}|}
\newcommand{\lfinm}[1]{\langle \phi^{-}_{#1}|}
\newcommand{\lfinn}[1]{\langle\phi^{\pm}_{#1}|}
\newcommand{\z}{\cal{Z}}
\newcommand{\RI}{\hat{{\cal{R}}}_{0}}
\newcommand{\Rt}{\hat{{\cal{R}}}_{\tau}}
\newcommand{\cb}{\bar{c}}
\newcommand{\nb}{\bar{n}}
\newcommand{\dnz}{ \delta n(\vec{r}_0,t)}
\newcommand{\dn}{ \delta n(t)}
\newcommand{\km}{ \kappa_{-}}
\newcommand{\dc}{ \delta c(\vec{x},t)}
\newcommand{\dcz}{ \delta c(\vec{x}_0,t)}
\newcommand{\dcw}{ \delta \hat{c}(\vec{x}_0,\omega)}
\newcommand{\dch}{ \delta \hat{c}(\vec{q},\omega)}
\newcommand{\dxi}{ \hat{\xi}_c(\vec{q},\omega)}
\newcommand{\dxib}{ \hat{\xi}_c(\omega)}
\newcommand{\dnh}{ \delta \hat{n}(\omega)}
\newcommand{\dnhq}{ \delta \hat{n}(\vec{q},\omega)}
\newcommand{\dnhz}{ \delta \hat{n}(\vec{r}_0,\omega)}
\newcommand{\dchz}{ \delta \hat{c}(\vec{r}_0,z_0,\omega)}
\newcommand{\nv}{  n(\vec{r},t)}
\newcommand{\cv}{ c(\vec{r},t)}
\newcommand{\nn}{\nonumber}
\newcommand{\rnb}{(\rho_0-\nb)}
\begin{abstract}
Many types of cells can sense external ligand concentrations with cell-surface receptors at extremely high accuracy. Interestingly, ligand-bound receptors are often internalized, a process also known as receptor-mediated endocytosis. While internalization is involved in a vast number of  important functions for the life of a cell, it was recently also suggested to increase the accuracy of sensing ligand as the overcounting of the same ligand  molecules is reduced.
Here we show, by extending simple ligand-receptor models to out-of-equilibrium thermodynamics, that internalization increases the accuracy with which cells can measure ligand concentrations in the external environment. 
Comparison with experimental rates of real receptors demonstrates that our model has indeed biological significance.
\end{abstract}

\maketitle
\section{Introduction}

Biological cells can sense and respond to various chemicals in their environment. However, the precision with which  a cell can measure the concentration of a specific ligand is  negatively affected by many sources of noise \cite{noise1,noise2,noise3,noise4,swain}.
 Most noticeable is external noise from the random arrival of ligand molecules at the cell-surface receptors by diffusion.
Nonetheless several examples exist in which
 measurements are performed with surprisingly high  accuracy.
In bacterial chemotaxis, for instance, fast moving bacteria such as {\it Escherichia coli} can respond to changes in concentration as low as 3.2 nM \cite{ecoli1}.
This value is remarkable, since cells have only about one second between ``tumbles'' 
to evaluate the ligand concentration \cite{bergbook}. Furthermore, this concentration value corresponds to only about three ligand molecules in the volume of the cell, assumed to be one femtoliter, suggesting single molecule detection. High accuracy is observed also in spatial sensing by single cell  eukaryotic organisms. Best characterized are  the slime mold {\it Dictyostelium discoideum},  which is able to sense a concentration difference of $1-5\% $ across the cell diameter \cite{dicty1},  as well as {\it Saccharomyces cerevisiae} (budding yeast), able to orient growth in a gradient of $\alpha$-pheromone mating factor down to estimated $1\%$ receptor occupancy difference across the cell \cite{segall93}.
Spatial sensing is also efficiently performed  by lymphocytes, neutrophils and other cells of the immune system
 \cite{zigmond}, as well as by growing synaptic  and tumor cells.
\begin{figure}[t]
\includegraphics[height=4.1 cm]{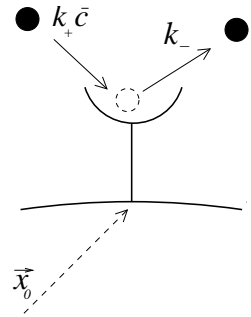}
%
\vspace{-0.05cm}
\caption{Single receptor, immobile at position $\vec{x}_0$,  binds and unbinds ligand  with rates $k_+ \cb$ and 
$k_-$, respectively.}
\label{fig_00}
\end{figure}

Previously,  the fundamental physical limits to the accuracy of sensing as set by ligand diffusion have been calculated \cite{purcell,bialek,levine1,levine2,levine3}. Recent work based on simplifying models indicates  that, if a cell effectively acts as an absorber of ligand,
the accuracy is significantly increased \cite{robned08}. Such an increase in accuracy can be explained with the fact that absorption prevents ligand molecules from unbinding the receptors. Hence, the same ligand molecule can only be  counted  once by a receptor, avoiding  a source of measurement uncertainty. However,  whether  cells with realistic receptors can act as absorbers and increase the accuracy of sensing is unknown.

 Motivated by these observations, in this paper we analyze the role of receptor-mediated endocytosis,
{\it i.e.}  the internalization of either bound or unbound receptors from  the cell membrane into the cell interior, often observed in eukaryotic cells \cite{muk,fergu}.
Internalization of ligand-bound receptors effectively leads to the absorption of ligand molecules
and is therefore expected to draw the cell nearer to the physical
limit  of the {\it perfect absorber} \cite{robned08}. 
Using simple models for the ligand-receptor dynamics,
 we find that the effect of receptor-mediated endocytosis
indeed increases the accuracy of sensing ligand concentration, if  internalization of ligand competes with ligand unbinding. Comparison of our results to the available literature of experimental rate constants
shows that receptors often work in this limit, indicating biological relevance of our results.
 
The paper is organized as follows:
In Section II we review the results regarding the accuracy of sensing
for a single,  immobile receptor without internalization.
In Section III,  we  study the role of 
internalization using  a model of ligand-receptor dynamics with internalization. While  ligand-receptor binding and unbinding is described by equilibrium thermodynamics,
 as previously developed in Refs. \cite{bialek,rob}, internalization clearly  introduces
 non-equilibrium thermodynamics into the problem. 
 We consider the limit
of fast diffusion, {\it i.e.} when
the coupling to  diffusion of ligand  can be neglected, as well as the general case both near and far from equilibrium. This general case deals with the reduction of rebinding of previously bound ligand molecules  by receptor internalization, and hence represents the main result of the work.
In Section IV we analyze the results obtained in Section III and
  connect to receptors from the  biological literature. We conclude with final comments
and discussion.  An Appendix is devoted to an alternative approach leading to the same results 
derived in the main text.
%

\section{Review of the single receptor}
 In this section we  review previous results for a single, immobile receptor without internalization. 
Details of the method will be provided in Section III.
 As  depicted in Fig. 1, such a  receptor can  bind and release ligand with rates $k_+ \cb$ and 
$k_-$, respectively. The kinetics for the occupancy $n(t)$ of the receptor are therefore given by
\BEQ
\label{basicsingle}
\frac{\partial n(t)}{\partial t}=k_{+} \bar{c}\left[1-n(t)\right] -k_- n(t),
\EEQ
where   the concentration of ligand $c(\vec{x},t)=\cb$ is assumed uniform and constant.
The steady-state solution for the receptor occupancy is given by
\BEQ
\label{nbKd}
\nb=\frac{\cb}{\cb+K_D}
\EEQ
with $K_D=k_-/k_+$ the ligand dissociation constant.
The rates of binding and unbinding are related to the (negative) free energy
of binding through detailed balance
\BEQ
\frac{k_+ \cb}{k_-}=e^{\frac{F}{T}}
\EEQ
with $T$ the temperature in energy units.
In  the limit of very fast ligand diffusion, {\it i.e.} when
a ligand molecule is immediately removed from the receptor after unbinding,
the dynamics of the receptor is effectively decoupled from
the diffusion  of ligand molecules and hence, diffusion does not need to be included explicitly.

Following Bialek and Setayeshgar \cite{bialek}, the accuracy of sensing is obtained  by applying the Fluctuation Dissipation Theorem (FDT) \cite{kubo}, which relates
the spectrum of the fluctuations in occupancy to the linear response  to a perturbation in the 
receptor binding energy. Furthermore  at equilibrium 
the fluctuations in occupancy can be directly related to the uncertainty in ligand concentration using Eq. (\ref{nbKd}).  For a measurement performed on a time scale $\tau$  much larger than the correlation time of the binding and unbinding events, the fluctuations of the occupancy $\langle (\delta n)^2\rangle$ are obtained from the zero-frequency spectrum divided by $\tau$.  Using  $\langle (\delta n)^2\rangle$,  one then obtains  the uncertainty in measuring ligand concentration $\cb$ \cite{bialek,rob}
\BEQ
\label{acc1}
\frac{\langle (\delta c)^2\rangle_{\tau}}{\cb^2}=  \frac{2}{k_+ \cb (1-\nb) \tau} \to \frac{1}{2 \pi D_3 \cb s \tau},
\EEQ
where 
 the right-hand side is obtained for diffusion-limited binding \cite{rob}, {\it i.e.} when $k_+ \cb (1-\nb) \to 4 \pi \cb  D_3 s$, with $D_3$ the diffusion constant and $s$ the dimension of the (spherical) receptor.
Eq. (\ref{acc1}) shows that the accuracy of sensing is limited by the random binding and unbinding of ligand.

\begin{figure}[t]
\includegraphics[width=0.7  \columnwidth]{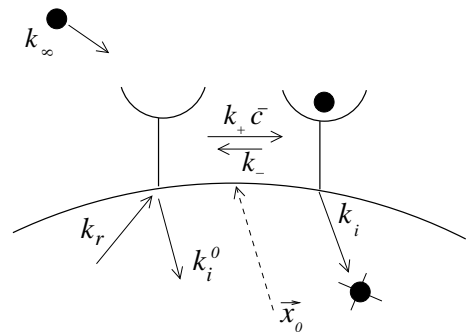}
\caption{ Receptor with internalization. In  addition to ligand binding and unbinding described in Fig. 1, the receptor is internalized at rate $k_i$, if occupied, and at rate $k_i^0$, if unoccupied. The  internalized ligand is degraded.  An unoccupied receptor is delivered to membrane with rate $k_r$.}
\label{potfig1}
\end{figure}

In  the case where  diffusion of ligand
is slow,   ligand binding to the receptor is affected by diffusion  \cite{bialek}.
The kinetics of the receptor occupancy and ligand concentration are 
described by
\begin{subequations}
\begin{align}
&\frac{\partial n(t)}{\partial t}=k_{+} c(\vec{x}_0,t)[1- n(t)] -k_-n(t) \label{2Dcoupleda}\\
 &\frac{\partial c(\vec{x},t)}{\partial t}=D_3 \nabla^2 c(\vec{x},t)-\delta(\vec{x}-\vec{x}_0)\frac{\partial n(t)}{\partial t} \label{2Dcoupledb},
\end{align}
\end{subequations}
where
 $\vec{x}_0$  indicates the position of the receptor and $\delta(\vec{x}-\vec{x}_0)$ is a Dirac delta function centered at the receptor location.  The last term in the second equation 
describes a sink or source of  ligand
at $\vec{x}_0$,
corresponding to ligand-receptor binding or unbinding, respectively. Analogous to fast diffusion, Eq. (5) has steady-state solutions $\cb$ (independent of $D_3$) and $\nb$
given by Eq. (2).

Following a similar procedure as in the previous case,
 the accuracy of sensing is given by \cite{bialek,rob}
\begin{subequations}
\label{dc3D}
\begin{align}
\frac{\langle (\delta c)^2\rangle_{\tau}}{\cb^2}
&= \frac{2 }{k_+ \cb (1-\nb) \tau} +\frac{1}{ \pi s D_3 \cb \tau} \label{dc3Da} \\
&\to \frac{3}{ 2 \pi s D_3 \cb \tau}, \label{dc3Db}
\end{align}
\end{subequations}
where the first term on the right-hand side of Eq. (\ref{dc3Da}) is the same as in Eq. (\ref{acc1}),
while the second term is 
the increase in uncertainty due to diffusion.
This  term accounts for the additional  measurement uncertainty from rebinding of  previously bound ligand to the receptor.
For diffusion-limited binding, one obtains  Eq. (\ref{dc3Db}) \cite{rob}.

Comparison of Eqs. (\ref{acc1}) and (\ref{dc3D})
 shows that removal of previously bound ligand by  fast  diffusion
increases the accuracy of sensing, since the same ligand molecule is never measured more than once. 

\section{Effect of receptor internalization}

\begin{figure}[t]
\includegraphics[width=0.7  \columnwidth]{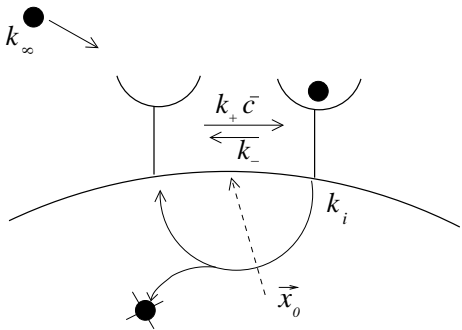}
\caption{ Single receptor with internalization. Similar to Fig. \ref{potfig1}, except that the delivery of receptor to the membrane is   adjusted, so that  Eq. (\ref{norm}) is fulfilled. This corresponds to an instantaneous replacement (round arrow) of the internalized  receptor by a new receptor via the endocytosis machinery. }
\label{potfig2}
\end{figure}

Here, we  consider the case of receptor internalization. 
As depicted in Fig. 2, receptors  at $\vec{x}=\vec{x}_0$ can
 bind and unbind ligand  with given rates. Furthermore, a bound receptorStren
can be   internalized at rate  $k_i$, while an unbound receptor can be internalized at rate  $k^0_i$.
Hence, the kinetics of the fractions of occupied receptors $n(t)$ and unoccupied receptors $m(t)$  are given by
\BES
\label{nmdyn}
\begin{align}
\frac{\partial n(t)}{\partial t}&=k_{+} \cb \, m(t) -(k_-+k_i)n(t)\label{nmdyna}\\
 \frac{\partial m(t)}{\partial t}&=-k_{+} \cb \, m(t) -k_i^0 m(t) +k_- n(t) +k_r. \label{nmdynb}
\end{align}
\EES
Imposing a single receptor at 
 $\vec{x}=\vec{x}_0$
 at any time  via
\BEQ
\label{norm}
n(t)+m(t)=1,
\EEQ
 Eq. (\ref{nmdynb})  becomes redundant.
As shown in Fig. \ref{potfig2}, this condition implies that an internalized receptor is immediately replaced by a new, unoccupied receptor with rate
$k_r(t)=k_i^0 m(t) +k_i n(t)$.
%
Furthermore,  rate  $k_{\infty}$ of incoming ligand  compensates for internalized ligand.

\subsection{Limit of fast diffusion}
We first  consider the case of fast diffusion, {\it i.e.} when ligand unbound from the receptor is immediately removed. 
In this case the kinetics  for
the occupancy $n(t)$ of the single receptor is  described by
 \BEA
\label{Int}
\nn \frac{\partial n(t)}{\partial t}&=&k_{+} \cb [1- n(t)] -(k_-+k_i)n(t)
\\
 &=&k_{+}\cb  [1- n(t)] -\kappa_- n(t), 
\EEA
where $\kappa_-=k_-+ k_i$ is the combined rate of unbinding and internalization.
The steady-state solution for the receptor occupancy is given by
\BEQ
\label{nbarInt}
\nb_{\it i}=\frac{\cb}{\cb+\kappa_-/k_+}=\frac{\cb}{\cb+K_M},
\EEQ
where $K_M=\kappa_-/k_+$ is a Michael-Menten-type  constant, and  subscript {\it i}  is used to  indicate
the  steady-state value for the occupancy of the receptor  in presence of  internalization ({\it cf}. Eq. (2)).
While Eq. (9) could be solved immediately by analogy to Eq. (1), we adopt here, for the fast diffusion case,  the  method of the effective temperature, which allows us to solve the general case in section III B.
Similar to the equilibrium case, at the non-equilibrium steady state
the rates can formally be  related to the binding free energy
\BEQ
\label{effect1}
\frac{k_+ \cb}{\kappa_-}=\frac{k_+ \cb}{k_- (1+\frac{k_i}{k_-})}=\frac{e^{F/T}}{1+k_i/k_-}=
e^{F/T_{e}},
\EEQ
 where we introduced 
the effective temperature
\BEQ
\label{Te}
T_e=\frac{T}{1- \frac{\ln(1+k_i/k_-)}{\ln{(k_+ \cb/k_-)}}}.
\EEQ
Hence, the effective temperature  maps the non-equilibrium steady state to an effective equilibrium,  allowing the generalization of the FDT to out-of-equilibrium phenomena \cite{cuglia,teo,crisanti,luca}  with  applications in modeling  biological processes \cite{lu}.
Conceptually, an effective temperature larger than the environment temperature ($T_e > T$) corresponds to a decrease in the receptor occupancy, approximately reflecting internalization in the non-equilibrium steady state.


In order to calculate the spectrum of the fluctuations in receptor occupancy,
 we follow Refs. \cite{bialek,rob} and consider 
small fluctuations
around the stationary solution
\BEQ
\nonumber n(t)=\nb_i+\delta n(t).
\EEQ
In order to apply the generalized  FDT (gFDT), we introduce fluctuations in the conjugate variable of the receptor occupancy, {\it i.e.} the free energy $F$, via fluctuations of the binding and unbinding rates 
\BEQ
\label{effect2}
 \frac{\delta F}{T_{e}}=\frac{\delta k_+}{k_+}-\frac{\delta \kappa_-}{\kappa_-},
\EEQ
where we approximate $T_e$ as a  parameter.

Linearization of Eq. (\ref{Int}) leads to

\BEQ
\label{deltanFInt}
\frac{ \partial [\dn]}{\partial t}=  -(k_+ \bar{c}\,+\kappa_-) \delta n(t) + k_+ \cb (1-\nb_i) \frac{\delta F(t)}{T_{e}},
 \EEQ
where we used Eq. (\ref{effect2}) to replace the fluctuations in the rate constants with fluctuations in the free energy, as well as  
 steady state solution  Eq. (10).

Fourier Transforming  Eq. (\ref{deltanFInt}) yields the susceptibility
\BEQ
\hat{\chi}(\omega)=\frac{\delta \hat{n}(\omega)}{\delta \hat{F}(\omega)}=\frac{1}{T_{e}} \frac{k_+ \cb (1-\nb_i)}{(k_+ \bar{c}\,+\kappa_-)-i\omega},
 \EEQ
describing the linear response of the receptor occupancy to a perturbation in the free energy.
We now use the  gFDT to calculate 
 the spectrum $S_n(\omega)=\langle |\delta n(\omega)|^2 \rangle$ of the fluctuations in $n(t)$
\BEA
\label{spectrumSRInt}
\nn S_n(\omega)&=& \frac{2 T_{e}}{\omega }
 \mbox{Im}[\hat{\chi}(\omega)]
=\frac{2 k_+ \cb (1-\nb_i)}{(k_+ \bar{c}\,+\kappa_-)^2+\omega^2 }=\\
 &=&2 \langle (\delta n)^2\rangle \frac{\tau_C}{1+(\omega\tau_C)^2},
\EEA
where  the correlation time $\tau_C=(k_+ \cb +\kappa_-)^{-1}$ 
and the total variance
\BEQ
\langle (\delta n)^2\rangle=\int_{-\infty}^{+\infty} \frac{d \omega}{2 \pi}S_n(\omega)=\frac{k_+ \cb (1-\nb_i)}{k_+ \cb+ \kappa_-}=\nb_i (1-\nb_i)\\
\EEQ
 have been introduced. 
Using Eq. (\ref{nbarInt}), we calculate the uncertainty in ligand concentration from fluctuations in occupancy
\BEQ
\label{km}
 \delta c=\frac{(\cb+K_M)^2}{K_M} \delta n=\frac{K_M}{(1-\nb_i)^2} \delta n=\frac{\cb}{\nb_i(1-\nb_i)}\delta n.
\EEQ
From Eq. (\ref{km}) the normalized  variance can be obtained
\BEQ
\label{nvar}
\frac{\langle (\delta c)^2\rangle}{\cb^2}= \frac{1 }{\nb_i (1-\nb_i)},
\EEQ
corresponding to an instantaneous measurement.

In the more realistic case, in which a measurement is performed
during an averaging time $\tau \gg \tau_C$, the error in the occupancy is linked to the low frequency spectrum via 
\BEQ
\label{szero}
\langle (\delta n)^2 \rangle_{\tau}\simeq \frac{S_n(0)}{\tau}=\frac{2 \nb_i^2 (1-\nb_i)}{k_+ \cb \tau}.
\EEQ
Using  Eqs. (\ref{km}) and (\ref{szero}), the accuracy of sensing is given by
\begin{align}
\frac{\langle (\delta c)^2\rangle_{\tau}}{\cb^2}=  \frac{2}{k_+ \cb (1-\nb_i) \tau} \label{avg}
 \to \frac{1}{2 \pi D_3 \cb s \tau}. 
\end{align}
Eq. (\ref{avg})  is identical to the result in Eq. (\ref{acc1}) without internalization, except that $\nb_i <\nb$ due to internalization.
In fact, the removal of unbound ligand by fast diffusion at equilibrium is equivalent to removal of bound ligand by internalization at the non-equilibrium steady state. This equivalence can be readily  seen from Eq. (\ref{Int}), which is indistinguishable
from simple unbinding with rate $\kappa_-=k_-+k_i$.
Hence, the effective temperature applied here is in fact exact.

\subsection{Solution near equilibrium}

When considering ligand diffusion, the above procedure
still applies
 with the exception that the concentration of ligand is allowed to vary due to binding and unbinding.
The kinetics of the receptor occupancy and ligand concentration is described by
\begin{subequations}
\label{2DInt}
\begin{align}
\frac{\partial n(t)}{\partial t}&=k_{+} c(\vec{x}_0,t)[1- n(t)] -\kappa_- n(t)
\label{2DInta}
\\
\frac{\partial c(\vec{x},t)}{\partial t}&=D_3 \nabla^2 c(\vec{x},t)-\delta(\vec{x}-\vec{x}_0)\left[\frac{\partial n(t)}{\partial t}+k_i n(t)\right] \notag\\
& +k_{\infty} \delta(\vec{x}-\vec{x}_{\infty}), 
\label{2DIntb}
\end{align}
\end{subequations}
where   $\kappa_-=k_-+k_i$ is used as before. Furthermore, a source of ligand 
 with rate $k_{\infty}$ is considered at location $\vec{x}_{\infty}$ 
 so as to compensate the loss of ligand molecules due to internalization.

We note that the steady-state solution for the concentration is not spatially uniform, but is depleted near the receptor due to internalization. This leads to the anomaly that we mathematically evaluate the rate of binding using the ligand concentration at $\vec{x}_0$ in Eq. (\ref{2DInta}), while physically
the diffusive flux and hence binding of ligand is determined by the ligand concentration $\cb$ far away from the receptor (see section III C). This is remedied by linearizing the ligand concentration around $\cb$ in the following.
Furthermore,  Eq. (\ref{norm}) is again assumed valid, and therefore an additional equation describing the unoccupied receptor fraction  $m(t)$ with rates $ k^0_i$ and $k_r$ is
redundant.

Linearizing Eqs. (\ref{2DInta}) and (\ref{2DIntb}) leads to 
\BES
\begin{flalign}
 \nn \frac{ \partial[ \dn]}{\partial t}&=  k_{+}(1- \nb_i) \delta c(\vec{x}_0,t) -(k_+ \bar{c}\,+\kappa_-) \delta n(t)
\\ &
+\delta k_+(t) \cb (1-\nb_i) -\nb_i \delta \kappa_-  \label{lineara}\\
 \nn \frac{\partial [\dc] }{\partial t}&=D_3 \nabla^2 \dc -\delta(\vec{x}-\vec{x}_0)\left[\frac{\partial [\dn]}{\partial t}
 \right. \\  & \left.
+ k_i 
\dn \rule{0 cm} {0.5 cm}
\right].\label{linearb}
\end{flalign}
\EES
By  applying the quasi-equilibrium picture with the effective temperature $T_e$  introduced in the previous subsection,
we use  Eq. (\ref{effect2}) to introduce fluctuations in the free energy, and obtain
\begin{widetext}
\BES
\label{linear2}
\begin{flalign}
  \frac{ \partial [\dn]}{\partial t}&=  k_{+}(1- \nb_i) \delta c(\vec{x}_0,t) -(k_+ \bar{c}\,+\kappa_-) \delta n(t)+ \kappa_- \nb_i \frac{\delta F}{T_{e}}
\label{linear2a}\\
 \frac{\partial [\dc] }{\partial t}&=D_3 \nabla^2 \dc -\delta(\vec{x}-\vec{x}_0)\left[\frac{\partial [\dn]}{\partial t} + k_i \dn  \rule{0 cm} {0.5 cm} \right]. \label{linear2b}
\end{flalign}
\EES
\end{widetext}
Fourier Transforming   Eqs. (\ref{linear2a}) and (\ref{linear2b}), we obtain
\BEQ
\label{ctrans}
\dch=e^{i \vec{q} \cdot \vec{x}_0} \frac{i \omega -k_i}{D_3 q^2 -i \omega}\dnh,
\EEQ
which can be inverse-Fourier Transformed  in $\vec{x}_0$ 
\BEQ
\label{antictrans}
\dcw=(i\omega-k_i)\dnh \int  \frac{d^3 q}{(2 \pi)^3} \frac{1}{D_3 q^2 -i \omega}. \EEQ
Inserting Eq. (\ref{antictrans}) in the Fourier-Transformed Eq. (\ref{linear2a}), we obtain 
\BEQ
\label{response}
\hat{\chi}(\omega)=\frac{\dnh}{\delta F(\omega)}=\frac{1}{T_e}\frac{\kappa_- \nb_i}{ k_+ \cb +\km +(k_i-i \omega)\Sigma_1(\omega) -i \omega},
\EEQ
where $\Sigma_1(\omega)$ is given by
\begin{flalign}
\label{Sigma10}
\nn \Sigma_1(\omega)&= \int \frac{d^3 q}{(2 \pi)^3} \frac{k_+ (1-\nb_i)}{D_3 q^2 -i \omega}
 = \int_0^\Lambda \frac{d q}{2 \pi^2} \frac{q^2[k_+ (1-\nb_i)]}{D_3 q^2-i \omega}\\
&  \overset{\omega \to 0}{=}\frac{k_+ (1-\nb_i)\Lambda}{2 \pi^2 D_3}\simeq \frac{k_+ (1-\nb_i)}{2 \pi  D_3 s}.
\end{flalign}
Here $\Lambda\simeq \pi/s$ is a cut-off due to the finite size $s$ of the receptor, introduced to regularize the integral in Eq. (\ref{Sigma10}).
As before,
 we apply  the gFDT 
to derive the spectrum of the fluctuations $\dnh$
\begin{flalign}
\label{FDTg}
\nn S_n(\omega)&=\frac{2 T_{e}}{ \omega} \mbox{Im}[\chi(\omega)]\\
&=\frac{2  k_+ \cb (1-\nb_i)[1+ \Sigma_1(\omega)]}{[ k_i \Sigma_1(\omega) +k_+ \cb +\kappa_-]^2 +\omega^2(1+\Sigma_1(\omega))^2 },
\end{flalign}
where we used Eq. (\ref{response}) for the susceptibility.
In the realistic case, in which the measurement  is time averaged over duration $\tau$  
much larger then the correlation time of the fluctuations, the relevant part of the spectrum is the zero frequency limit
\BEA
\label{S0}
\nn S_n(\omega \simeq 0)&=& \frac{2  k_+ \cb (1-\nb_i)[1+ \Sigma_1(0)]}{[ k_+ \cb +\kappa_- + k_i \Sigma_1(0)]^2}\\
\nn &=&2  \langle (\delta n)^2\rangle \frac{(1+\Sigma_1(0))\tau_C}{(1+k_i \Sigma_1(0) \tau_C)^2}\\
 &\simeq& 2  \langle (\delta n)^2\rangle  \left[1+ \alpha \Sigma_1(0)\right] \tau_C,
\EEA
where $\alpha=1-2 k_i \tau_C <1$, and higher order terms in $\Sigma_1(0)$ are neglected for sufficiently fast diffusion.
As before,  $\langle (\delta n)^2\rangle=\nb_i(1-\nb_i)$ and $\tau_C=(k_+  \cb + \kappa_-)^{-1}$.

Using Eq. (\ref{km}), the normalized variance of the concentration is given by
\BEQ
\frac{\langle (\delta c)^2\rangle_{\tau} }{\cb^2}=\frac{\langle (\delta n)^2 \rangle_{\tau}}{[\nb_i(1-\nb_i)]^2} =\frac{1}{[\nb_i(1-\nb_i)]^2} \frac{S_n(0)}{\tau},
\EEQ
where $\tau$ is the averaging time.
Using Eq. (\ref{S0}) for the power spectrum, we finally obtain for the accuracy of sensing
with ligand  internalization and diffusion
\begin{subequations}
\label{dc3DINT}
\begin{align}
 \frac{\langle \left(\delta c\right)^2\rangle_{\tau} }{\cb^2}&=\frac{2}{k_+ \cb (1-\nb_i) \tau}+\frac{\alpha}{\pi D_3 \cb s \tau} \label{dc3DINTa}\\
&\to \frac{1+2 \alpha}{2 \pi D_3 \cb s \tau}.\label{dc3DINTb}
\end{align}
\end{subequations}

The following conclusions can be drawn by comparison with the result Eq. (\ref{dc3D})  without internalization:
({\it i}) Receptor internalization  mainly reduces the second term
in Eq.(\ref{dc3D}),
demonstrating for the first time that internalization increases the accuracy of sensing by reducing the uncertainty from rebinding of previously bound ligand.
The first term is only reduced by replacing
$\nb$ by $\nb_i$, with  $\nb_i<\nb$.
({\it ii}) In the limit $k_i \to 0$, $\alpha \to 1 $
 and the equilibrium result Eq. (\ref{dc3D})
without internalization   is recovered.
({\it iii}) As  Eq. (\ref{dc3DINT}) becomes unphysical  in the limit of $k_i \to \infty$, {\it i.e.} does not approach Eq. (\ref{avg}) without rebinding, our result can only be regarded   an approximation valid near equilibrium.

\subsection{Comparison with perfect absorber}

The perfect absorber is here defined as a receptor, which internalizes a ligand immediately once it is bound.
Following Ref. \cite{robned08}, the accuracy of sensing can be calculated from the Poisson statistics of the number of binding events $N$ in time $\tau$ 
\BEQ
\label{dcdN}
\frac{\langle(\delta c)^2\rangle_{\tau}}{\cb^2}=\frac{\langle(\delta N)^2\rangle}{\langle N^2\rangle}=\frac{1}{4 \pi D_3 s \cb \tau},
\EEQ
obtained from the diffusive flux of ligand to an absorbing sphere of radius $s$. Comparison with Eq. (\ref{avg}) for diffusion-limited binding  without rebinding shows that the perfect absorber is yet more accurate by a factor $2$. This is due to the fact that the fluctuations in occupancy in Eq. (\ref{avg}) stem from the random binding {\it and} unbinding/internalization events, while the uncertainty in Eq. (\ref{dcdN}) solely stems from the random binding events (see also factor 2 in Eq. (\ref{lang1xi}), as well as  Ref. \cite{robnedprl} for further explanation).

\section{Implications for Biology}

In the previous section we showed that internalization 
increases the accuracy of sensing by reducing the measurement uncertainty 
from rebinding of previously bound ligand. 
In this section we review some receptors of known rate constants.
We specifically would like to determine if the rate of internalization $k_i$ is fast enough, {\it i.e.} comparable to 
 the unbinding rate $k_-$, in order to effectively increase the
 accuracy of sensing.

In Table I,  we summarize experimental values for rate constants, including internalization, of various receptors.
Most G-protein coupled receptors (GPCR) undergo internalization \cite{fergu}. The  Ste2 receptor
 in haploid yeast cells of $\alpha$-mating type is involved in  $\alpha$ pheromone sensing
and signal transduction, leading to  cell polarization, `shmoo' formation and mating.
The folate receptor (FR) in {\it Dictyostelium}, likely  a GPCR \cite{rifkin, dewit2}, 
is used to sense and hunt bacteria. (The folate-binding protein in mammalian cells is a diagnostic marker for various cancers, and its internalization is exploited for drug delivery into cancerous human cell \cite{paulos}.)
However, 
 the cAR1 receptor in {\it Dictyostelium}, used for  sensing of cAMP under starvation, is not internalized \cite{catari}.
The epidermal growth factor receptor (EGFR), a tyrosine kinase, is another important example of a receptor which is internalized \cite{shanka}. This receptor is involved in cell growth, proliferation and differentiation \cite{shanka,bianco,rorth}. Another class of internalized receptors is involved in uptake.
 Transferrin receptor (TfR)  is  used for iron uptake from extracellular space
and plays therefore an important role in blood cells \cite{shanka}. Transferrin binds to  TfR, is  internalized and releases its iron load  through ion pump-induced pH reduction.
The ligand-bound TfR is then recycled back to cell surface.
Another example is the  low density lipoprotein  receptor (LDLR) \cite{shanka}. 
When bound to LDL-cholesterol via adaptin, 
LDLR is internalized via
clathrin-coated vesicles  \cite{shanka}.
Furthermore, the vitellogenin receptor (VtgR) is  involved in oogenesis (egg-formation) \cite{shanka}. Once internalized, vitellogenin is  turned into yolk proteins. Ligand-free receptors are recycled back to cell surface.
Table I shows 
 that
 in most cases $k_i$ is of the same order or larger than $k_-$, except
for FR, where  internalization  is much slower than
the unbinding of the ligand from the receptor.

\begin{widetext}
\begin{center}
\begin{table}[ht!]
\begin{tabular}{|l||l|l|l|l|l|}
\hline
 \;  Receptor & \;Function & $ \;k_-$ (min$^{-1}$) & $K_D$ (nM ) & $k_i$ (min$^{-1})$ & $k_i^0$ (min$^{-1}$)\\
\hline
\hline
 & {\it chemotaxis:}& & & &\\
FR  & feeding & 0.096 \cite{frazier} & 20.0 \cite{frazier}$^{\dagger}$  & 9.6$\times 10^{-4}$ \cite{paulos}$^{*}$& -- \\
Ste2   & mating & 0.06 \cite{raths}  &22.1 \cite{raths}  &  0.24 \cite{yi} & 0.024 \cite{yi} \\
  &        &      0.0108 \cite{jennes}   & 6.0 \cite{jennes}      &0.156 \cite{hicke}       &0.0156 \cite{hicke} \\
EGFR \cite{shanka} & development & 0.24 & 2.47 & 0.15 &0.02 \\
\hline 
& {\it uptake:} & & & &\\
TfR \cite{shanka}& iron  &0.09   & 29.8 & 0.6 & 0.6\\
LDLR \cite{shanka} & cholesterol  & 0.04  & 14.3  &0.195 &0.195 \\
VtgR\cite{shanka} &vitellogenin  &0.07   & 1300  & 0.108& 0.108\\
 & & & & & \\
\hline
\end{tabular}
\label{table1}
\caption{Summary of experimental data for relevant receptor rates discussed in the main
text. $^{\dagger}$ However, other values have been reported as well \cite{segall2}.$^{*}$ This rate is measured for folate-binding protein in cancerous  mice cells, not in {\it Dictyostelium}. }
\end{table}
\end{center}
\end{widetext}

Figure 4 visualizes the  contribution of internalization to the accuracy of sensing for the receptors from Table I.
 The faster the internalization, the larger the increase of the accuracy of sensing.
However internalization can only reduce the second term in Eq. (\ref{dc3Da}) from rebinding of previously bound ligand,  not the first term from random binding and unbinding. To illustrate the relative contribution of the two terms, we plot the square root of their ratio in Fig. 4 (filled bars). 
This shows that in most cases, in which internalization occurs,
the noise ratio of the two terms is significant, 
ranging
 from few hundredths to order of unity. Hence,
 internalization can lead to a substantial increase in the accuracy of sensing.
However, measured rate constants are substantially uncertain (see below) and  diffusion coefficients of small ligand molecules range from $0.1 \mu$m$^2$/s
 in the synaptic cleft between neurons \cite{syncleft} to $1-10 \mu$m$^2$/s in blood  \cite{blood} 
 to $300 \mu$m$^2$/s  in water \cite{robned08}.
In order to avoid this uncertainty in parameters, we also plot the upper  limit of the noise ratio for diffusion-limited binding  equal to $\sqrt{2}$ (dashed line in Fig. 4). Removal of the second noise term in this limit by internalization would increase the accuracy in Eq. (6b) by a factor 3.
In Fig. 4, we also show  the  strength of internalization, defined by the ratio of internalization and unbinding rates (open bars). 
\begin{figure}[tht]
    \includegraphics[width=1 \columnwidth]{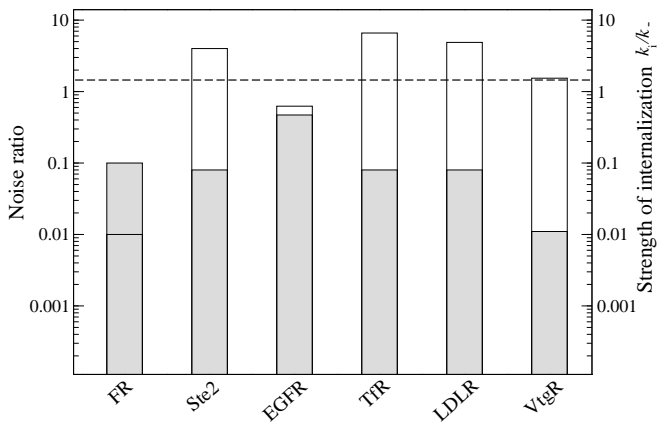}
   \label{test4}
 \caption{
Analysis of receptor rate data from Table I. (Filled bars)  Noise ratio defined as the square root of the second  (rebinding) term and the first (random binding and unbinding) in Eq. (6a). 
(Open bars)  Strength of internalization, defined by ratio  $k_i/k_-$.The dashed line indicates the noise ratio ($\sqrt{2}$) for diffusion-limited binding ($k_+\cb (1-\nb) \to 4 \pi \cb D_3 s$) . Numerical values of the plot are provided in Appendix B.}
\end{figure}

 How reliable are the measured values for the rate constants?
Rate constants are generally obtained through radioactive labeling of ligand, with the isotope
choice targeted to each specific case (the isotope $^{125}I$ giving the most accurate measurements) \cite{book}.
In order to measure the unbinding and the binding rates, {\it i.e.} $k_-$ and
$k_+=k_-/K_D$, the receptors on the membrane must be separated by filtration
or centrifugation from the soluble ligand. If the ligand-unbinding process is slow
compared to  separation, then the measurement of the amount of bound ligand
through the radioactive label can be easily carried out; in the case of fast unbinding, measurements are less accurate.
For the internalization rate,  the ratio between the intensity at the surface and inside the cell
is  measured and from the slope of the time variation of this ratio,
$k_i$  is determined \cite{radio}. This method, though, does not take into account recycling (diacytosis) of the receptor, which is a recurrent feature of the internalization process, also included in our model.
These shortcomings, as well as the
 variability associated   with different cell preparations,
lead to
a large error in the rate determination  ($\sim 20 $\%) \cite{book}
and variability between different measurements (20-90\%).
Other measurement methods employed in experiments include protease sensitivity assays \cite{schandel} and destination assays \cite{blumer}. 
 
 Many examples in fact suggest a direct relation between ligand internalization  and the accuracy
of sensing, measured by the sensitivity of  cell polarization or cell movement in shallow chemical gradients. 
These include sensing of $\alpha$-factor
 by budding  yeast 
\cite{hicke,jennes},  folate
by  {\it Dictyostelium} \cite{rifkin} and PDGF
 by fibroblasts \cite{kawada}.
Other examples relate to embryonic development. In zebrafish,  primordial  germ cells migrate toward chemokine SDF-1a that binds and activates the receptor CXCR4b. It was recently shown that ligand-induced CXCR4b internalization is required for precise arrival of germ cells at their
target destination \cite{minina}. During {\it Drosophila} oogenesis,
 border cells 
 perform directional migration \cite{bianco}.
  EGFR, together with two  other  receptor tyrosine kinases,
is the main guidance receptor. Recent work 
in this system provided compelling evidence  that guided  cell movement also requires spatial control of signaling events by endocytic dynamics
  \cite{luque,leroy}.
%

\section{Discussion and Conclusions}

In this paper we  analyzed the role of receptor
 internalization in the accuracy of sensing ligand concentration.
By extending equilibrium single receptor  models 
to  non-equilibrium  thermodynamics introduced by internalization, we derived expressions
for  the uncertainty in sensing ligand concentration.
As expected, internalization of  ligand-bound receptors   makes the cell act similarly to  an absorber and increases the accuracy of sensing.
We then analyzed relevant experimental data, summarized in Table I,
and concluded that in most cases, the contribution of receptor internalization  to the increase in accuracy of sensing
is non-negligible. However, a perfect absorber is yet more accurate as its uncertainty only stems from random binding events, not from additional random unbinding {\it and } internalization events. Whether cells have developed  mechanisms to approach the limit of the  perfect absorber, {\it e.g.} by internalization, is not clear yet. However, since the accuracy of concentration sensing can always be improved by increasing the averaging time, one might expect that receptor internalization becomes increasingly important when time is of the essence. In addition to chemotaxis, embryonic development may be a biological system for which receptor internalization is important. Specifically,  the Fgf8 morphogen, which regulates tissue differentiation and morphogenesis in zebrafish, forms exponential concentration gradients by diffusion and degradation, the latter being achieved precisely by receptor internalization \cite{fgf8}. However, based on our results,  internalization may also be used  for the accurate readout of the gradient in the short amount of time dictated by cell division.
 
Cells generally have many receptors to estimate external concentrations of chemicals, leading to a spatial averaging and consequently further increase in the  accuracy of sensing. However, even the employment of  many receptors cannot increase the accuracy of sensing beyond the physical limit of the perfect absorber\cite{ robnedprl}. In fact a cell only needs a relatively small number of receptors to achieve  an accuracy comparable to  the physical limit  \cite{bergbook}. On the downside, if a cell uses many receptors, it needs to integrate this information in signaling pathways. If this process is fundamentally limited by noise as well, then this noise provides an upper limit on the overall estimation performance.


Although
we analyzed the role of receptor internalization in increasing  the accuracy of sensing, it is important to stress  that 
internalization fulfills several other known  purposes in cells  \cite{ira}.
 Among them are
 ({\it i}) redistribution of receptors to different locations on the cell membrane,
({\it ii}) uptake of nutrients  and chemicals, ({\it iii}) signaling by ligand  in cell interior, and ({\it iv}) turning off persistent
signal as part of adaptation. All these aspects would need to be considered to fully characterize  the working of a receptor.

In order to derive the accuracy of sensing with internalization, we made a number of simplifying assumptions. 
In our model we  neglected other possible 
sources of fluctuations such as fluctuations  in receptor density, in order to relate our results to 
the single, immobile receptor.
Furthermore, we introduced  
the effective temperature $T_e$ to generalize the FDT to  non-equilibrium thermodynamics. While $T_{e}$ is well defined for well separated  time scales, a potential time or frequency dependence of $T_{e}$ \cite{teo,crisanti,luca} was neglected here. However, our non-equilibrium  result with internalization is consistent with the equilibrium  result  in the fast diffusion limit.  Removal of  ligand by fast diffusion at equilibrium is   equivalent to removal
of bound ligand by internalization at the non-equilibrium steady state, providing confidence in our method.

Internalization is not the only mechanism, by which a cell  can act as an  absorber and increase its accuracy of sensing. Other potential mechanisms include
enzymatic degradation of ligand at the cell surface, {\it e.g.}
degradation of cAMP  by  mPDE 
in {\it Dictyostelium} and  of $\alpha$-mating pheromone by Bar1  in budding yeast  \cite{hicks,barkai}. Furthermore, at excitatory neural synapses, fast diffusion of AMPA  receptors   on the  post-synaptic membrane surface
 has an important role in the sensing of  neurotransmitter  glutamate \cite{choquet}. 
Ligand-bound, desensitized receptors diffuse away and are replaced by fresh receptors, leading to fast recovery and readiness for the next action potential  and release of neurotransmitter. By the same mechanism, the accuracy of sensing may be increased, since ligand-bound receptors diffuse away and release ligand far away from region of signaling, thus preventing an overcounting of same ligand molecules.
\cite{pap2}.

\acknowledgments
We thank Robert Insall, Luca Leuzzi, Yigal Meir and Ned Wingreen  for helpful discussions, and two anonymous referees
for their helpful comments. We acknowledge financial support from
Biotechnological and Biological Sciences Research Council grant BB/G000131/1
and the Centre for Integrated Systems Biology at Imperial College.

\appendix

\section{Langevin approach}

As an alternative derivation, here we provide the solution for the accuracy of sensing for internalization using a Langevin approach.
We first consider fast diffusion. We start from  Eq. (\ref{Int}) but add a noise term, $\xi_n(t)$
\BEQ
\label{lang1}
\frac{\partial n(t)}{\partial t}=k_{+} \cb [1- n(t)] -\kappa_- n(t) +\xi_n(t),
\EEQ
where we assume
\BEQ
\label{lang1xi}
\langle |\hat{\xi}_n(\omega)|^2\rangle=k_+ \cb(1-\nb_i)+\kappa_- \nb_i=2 k_+ \cb(1-\nb_i)
\EEQ
due to Poisson statistics \cite{langevin1,langevin2}.
Linearizing and Fourier Transforming Eq. (\ref{lang1}), assuming the rates $k_+ $ and $\kappa_-$ constant, leads to
\BEQ
\label{dnlang1}
\delta \hat{n}(\omega)=\frac{\hat{\xi}_n(\omega)}{ k_+ \cb+\kappa_- - i \omega}.
\EEQ
Hence, the power spectrum of the fluctuations in receptor occupancy is given by
\BEA
\label{snlg}
\nn S_n(\omega)=\langle |\delta \hat{n}(\omega)|^2 \rangle &=&\frac{\langle |\hat{\xi}_n(\omega)|^2\rangle}{(k_+ \cb+\kappa_-)^2 +\omega^2}\\
&=& \frac{2 k_+ \cb(1-\nb_i)}{(k_+ \cb+\kappa_- )^2 +\omega^2},
\EEA
where in the last step the property  Eq.(\ref{lang1xi}) was used.
Eq. (\ref{snlg}) is indeed equivalent to result  Eq. (\ref{spectrumSRInt}) in the main text.

For the general solution, we  start  from the Fourier-Transformed 
Eq. (\ref{linear2}) and Eq. (\ref{antictrans}), {\it i.e.}
\BEQ
\label{sldif1a}
(k_+ \cb+ \kappa_-- i \omega) \delta \hat{n}(\omega)= k_+ (1-\nb_i) \dcw
+ \hat{\xi}_n(\omega)
\EEQ
and
\BEQ
 \delta \hat{c}(\vec{x}_0, \omega)= \frac{(i \omega -k_i)}{k_+ (1-\nb_i)}\Sigma_1(\omega) \dnh +\hat{\xi}_c(\omega) ,
\label{sldif1b}
\EEQ
respectively, where $\hat{\xi}_n(\omega)$ and $\hat{\xi}_c (\omega)$ are additive noise terms, and  
$\Sigma_1(\omega)$ is  given by Eq. (\ref{Sigma10}).
Inserting Eq. (\ref{sldif1b})  in Eq. (\ref{sldif1a}) and solving for $\dnh$ leads to
\BEQ
\label{dnlang2}
\dnh=\frac{\hat{\xi}_n(\omega) +k_+ (1-\nb_i)\hat{\xi}_c(\omega)}{k_+ \cb +\kappa_- +k_i \Sigma_1(\omega)- i\omega(1+\Sigma_1(\omega))},
\EEQ
from which   the following expression for $\langle |\delta n(\omega)|^2 \rangle$ ensues
\BEQ
\label{dnsqlang}
\langle |\delta \hat{n}(\omega)|^2 \rangle =\frac{k^2_+ (1-\nb_i)^2 \langle |\dxib|^2 \rangle +\langle |\hat{\xi}_n(\omega)|^2\rangle}{[k_+ \cb +\kappa_- + k_i \Sigma_1(\omega)]^2 + \omega ^2 [1+\Sigma_1(\omega)]^2}.
\EEQ 
In the limit $\omega \to 0$, using  Eq. (\ref{lang1xi})  as in the previous case, we obtain
\begin{flalign}
\label{dnsqlang2}
\langle |\delta \hat{n}(\omega)|^2 \rangle \overset{\omega \to 0}{=}\frac{k^2_+ (1-\nb_i)^2 \langle |\dxib|^2 \rangle
+ 2 k_+ \cb (1-\nb_i)}{[k_+ \cb +\kappa_- + k_i \Sigma_1(0)]^2}.
\end{flalign}
 Following \cite{bialek2}, we set
\BEQ
\label{bk2}
\langle |\dxib|^2 \rangle\simeq S_c^{3D}(\omega\to0)\sim \frac{\cb}{D_3 s}\sim \frac{\Sigma_1(0)}{k_+ (1-\nb_i)}
\EEQ
%
in Eq. (\ref{dnsqlang2}), and obtain for the power spectrum 
\begin{align}
\label{dn2lg}
\langle |\delta \hat{n}(\omega)|^2 \rangle = \frac{2\langle (\delta n)^2 \rangle  (1+\Sigma_1(0))\tau_C}{[1+k_i \Sigma_1(0) \tau_C]^2}
\end{align}
Eq. (\ref{dn2lg}) is identical to    result  Eq. (\ref{S0}), obtained with the gFDT in  main text.

\section{Numerical values}
In this section we provide the numerical values used for plotting Fig. 4.
Noise ratio and internalization strengths  
 are respectively given by:
0.1/0.01 (FR),  0.08/4 (Ste2), 0.47/0.625 (EGFR), 0.08/6.6 (TfR), 0.08/4.875 (LDLR), 0.011/1.54 (VtgR). Specifically, to calculate the noise ratio, the first and the second term in Eq. (6a) are given by (in units of $\tau$):
2500/27.6 (FR),  4000/25.1 (Ste2), 1000/224 (EGFR), 2667/18.6 (TfR), 6000/38.72 (LDLR),
3428/0.426 (VtgR). We have used $s=1nm$, $\nb=1/2$, {\it i. e.} setting $\cb=K_D$ from Table I, and
$D_3= 1 \; \mu $m$^2/$s .

\end{document}